\documentclass[12pt]{iopart}
\usepackage{graphics}
\usepackage{epsfig}
\begin{document}

\title{Superconducting qubit network with controllable nearest neighbor
coupling}
\author{M. Wallquist, J. Lantz, V.S. Shumeiko, and G. Wendin}
\address{Department of Microtechnology and Nanoscience,
Chalmers University of Technology, 41296 Gothenburg, Sweden.}
\ead{wendin@mc2.chalmers.se}

\begin{abstract}
We investigate the design and functionality of a network of loop-shaped charge
qubits with switchable nearest-neighbor coupling. The qubit coupling is
achieved by placing large Josephson junctions at the intersections of the
qubit loops and selectively applying bias currents. The network is
scalable and makes it possible to perform a universal set of quantum
gates. The coupling scheme allows gate operation at the charge degeneracy
point of each qubit, and also applies to charge-phase qubits.
Additional Josephson junctions included in the qubit loops for qubit
readout can also be employed for qubit coupling.
\end{abstract}


\maketitle
%
\section{Introduction}

During the last six years it has been experimentally proven that
superconducting circuits can serve as quantum mechanical two-level systems,
qubits, to be used for quantum information processing
\cite{Nakamura,Vion,vanderWal,Martinis,Chiorescu,Saclay2004}. Besides the
experiments with individual qubits, several experiments have been performed
so far on two permanently coupled qubits
\cite{Pashkin,Yamamoto,Berkley,Izmalkov,McDermott}. For instance, to observe
the coupling of two charge qubits, the qubit islands have been permanently
coupled via a capacitor, and the strength of the coupling has been varied by
tuning the qubits in and out of resonance with each other (by varying the
gate voltage) \cite{Pashkin,Yamamoto}.

In order to build a functional, scalable quantum computer,
a network design is needed
that allows coupling of an arbitrarily large number of qubits, with the
possibility to switch on and off the coupling by means of external
control knobs. In principle,  coupling of only nearest
neighbor qubits is sufficient to perform a universal set of gates
\cite{Schuch}.

Theoretical schemes for variable coupling of charge qubits have been
intensely discussed in literature. Couplings via inductive and
capacitive elements have been examined as well couplings via linear
LC-oscillators and Josephson junctions
\cite{Shnirman,Makhlin,Plastina,Siewert,YouPRL}. A standard approach to
achieve a variable coupling is to employ a SQUID-type geometry either for
the qubits \cite{MakhlinRMP} or for the coupling element \cite{Blais}, to
be able to control the Josephson energy by an external magnetic flux. A
somewhat different approach has been suggested in \cite{AverinBruder},
where the qubits are coupled via another charge qubit thus creating a
variable capacitive coupling.

Recently, a different way to achieve a variable inductive coupling has been
suggested, namely to let the charge-qubit loops
intersect and share a coupling Josephson junction or SQUID. The
interaction is then controlled either by varying the magnetic flux in the
qubit loops (or the coupling SQUID) \cite{YouPRB}, or by applying bias
currents to the coupling Josephson junction \cite{Lantz}.

In this paper we give a detailed analysis of the qubit network based on
the coupling method proposed in \cite{Lantz}. The idea of this method is
to couple loop-shaped charge qubits by letting the circulating loop currents,
which are sensitive to the charge state of the qubit
island, interact. This is done by placing a non-linear oscillator - a
large Josephson junction - at the intersection of the qubit loops. Such a
coupling can be made variable by using the fact that in the absence of an
external magnetic field, the persistent currents in the qubit loops are
absent (for symmetric qubits with equal Josephson junctions). However,
when a dc current bias is applied to a coupling Josephson junction the
symmetry is broken and currents start to circulate, the magnitude of
the currents being dependent on the bias current. These currents interact
with the oscillator, resulting in a variable effective qubit-qubit
coupling. A similar coupling effect can be accomplished by inserting large
readout Josephson junctions in the qubit loops
 \cite{Vion}. As we show in this paper, applying current to one of the
readout junctions allows one to measure the state of the
corresponding qubit without disturbing the other qubits; however, when two
neighboring readout junctions are biased, the qubit-qubit coupling is
switched on.

The advantage of the current-biased coupling scheme is that it does not
need any local
magnetic fields to control the coupling, fields which could create unwanted
parasitic long range interactions. An important feature is the possibility to
operate at the qubit charge degeneracy point, where the decoherence effect
is minimized \cite{Vion,Duty}, and where the gate operations are
very simple.
This coupling scheme can be also extended to charge-phase qubits which
are still less sensitive against decoherence due to flatter band structure
\cite{Vion}. With this
coupling mechanism, neighboring qubits in an arbitrarily
long qubit chain can be
coupled, and several independent two-qubit gates can be performed
simultaneously. The fundamental entangling two-qubit gate is a
control-phase gate, which together with single-qubit gates constitute a
universal set of operations.

The structure of the paper is as follows: In section \ref{s2q} we explain
the principles of the coupling by considering the simplest case of two
coupled qubits. We estimate the maximum coupling strength, evaluate the
residual parasitic couplings, and investigate the charge-phase regime for
the qubits. In section \ref{s2qm} we add measurement junctions to the
qubit circuits and investigate how to use these junctions as read-out
devices and the means to create qubit coupling. In section \ref{sNq} we
generalize the derivation of section \ref{s2qm} to a multi-qubit network
with an arbitrary number of coupled qubits. Finally, in section
\ref{suse}, we discuss how to use the network for quantum computing.

\section{Controllable coupling of two qubits}
\label{s2q}
\begin{figure}[hbt]
\begin{center}
\includegraphics[width=5cm]{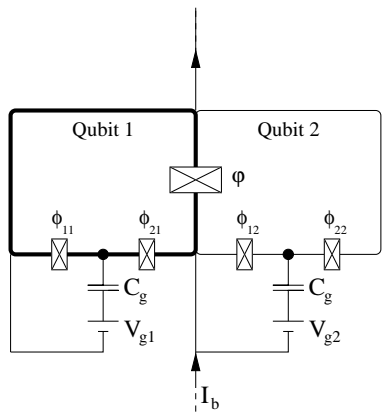}
\caption{A system of two coupled charge qubits. $V_{gi}$ control the
individual qubits whereas $I_{b}$ controls the coupling of the two
qubits.} \label{2qub}
\end{center}
\end{figure}

To more clearly explain the principle of the qubit coupling, we first
consider the case of two coupled qubits. The qubits consist of single Cooper
pair boxes (SCB)
with loop-shaped electrodes.
\cite{Vion,Zorin}. To create coupling between the qubits,
a large-capacitance
Josephson junction (JJ)
is placed at the intersection of the qubit loops, see Fig. \ref{2qub}. The
physics of the coupling is the following: As long as no magnetic flux is
applied to the qubit loops, and no bias current is sent through the coupling
junction, there are no circulating currents in the qubit loops. However, when
the bias current is switched on, circulating currents start to flow in the
 clockwise or counterclockwise direction depending on the charge state of
the Cooper pair box. These currents displace the coupling junction oscillator
and change its ground state energy, leading to an effective qubit-qubit
interaction. The strength of the interaction is proportional to the bias
current through the coupling junction. In the idle state, when the bias
current is switched off, small phase fluctuations at the coupling junction
generate permanent parasitic qubit-qubit coupling. However, this parasitic
coupling can be made small compared to the controllable coupling by choosing
the plasma frequency  of the coupling junction $\omega_b=\sqrt{2E_J^bE_C^b}$
to be small compared to the Josephson energy $E_J^b$. This requirement
implies that the junction charging energy $E_C^b$ must be small,
\begin{equation}\label{EjEc}
E_C^b \ll \omega_b \ll E_J^b \, ,
\end{equation}
i.e. the coupling junction must be in the phase regime. This is the most
essential requirement for the qubit coupling under consideration.

\subsection{Circuit Hamiltonian}
\label{ss2qL}
We begin the evaluation of qubit coupling with the derivation of a circuit
Hamiltonian. To this end, we first write down the Lagrangian $L$ of the
circuit in Fig. \ref{2qub}. The Lagrangian consists of the respective
Lagrangians of the SCBs and the coupling JJ,
\begin{equation}
L = \sum^2_{i=1} L_{SCB,i}+L_{JJ} \,.
\label{Lorig}
\end{equation}
Assuming the single Cooper pair boxes to consist of identical junctions with
capacitance $C$ and Josephson energy $E_J$, and following the rules described
in e.g. Refs. \cite{Devoret,Yurke}, we write the corresponding Lagrangian on
the form,
\begin{equation}\label{LSCB}
\fl
L_{SCB,i} = \frac{\hbar^2C}{2(2e)^2}\left(\dot{\phi}_{1i}^2 +
\dot{\phi}_{2i}^2
\right)
+ \frac{\hbar^2C_{g}}{2(2e)^2}\left(\frac{2e}{\hbar}V_{gi}-\dot{\phi}_{1i}
\right)^2
+  E_{J}\left(\cos \phi_{1i} + \cos \phi_{2i}\right),
\end{equation}
where $\phi_{1i}$ ($\phi_{2i}$) is the phase difference across the left
(right) Josephson junction of the $i$-th SCB, and $C_g$ is the gate
capacitance. The Lagrangian of the
coupling JJ includes the electrostatic energy and the Josephson
energy of the junction, and also the interaction energy of the junction with
applied bias current $I_b$,
\begin{equation}\label{LJJ}
L_{JJ} = \frac{\hbar^2C_b}{2(2e)^2}\dot{\varphi}^{2} + E_{J}^b\cos \varphi +
\frac{\hbar}{2e}I_{b}\varphi,
\end{equation}
where $\varphi$ is the phase difference across the coupling junction.

The flux quantization condition in each of the qubit loops allows the
elimination of one of the qubit variables from the Lagrangian. We assume that
there is no external magnetic flux in the loops since magnetic flux will not
be used to control either the qubits or the qubit interaction, and we also
assume that the loop self-inductances are negligible. Then the flux
quantization equation takes the form,
\begin{equation}\label{phases}
\phi_{+,1}+\varphi=0,\qquad \phi_{+,2}-\varphi=0\,.
\end{equation}
where we introduced new qubit variables,
\begin{equation}\label{phi_pm}
\phi_{-,i}=\frac{\phi_{2i}-\phi_{1i}}{2},\qquad \phi_{+,i}=\phi_{1i}+\phi_{2i}.
\end{equation}
By virtue of relations (\ref{phases}), the gate capacitance terms will take
the form
$(\hbar^2C_g/2(2e)^2)(2eV_{gi}/\hbar+\dot{\phi}_{-,i} \pm \dot{\varphi}/2)^2$
introducing a capacitive interaction of the SCB with the coupling junction.
From
here on, the
upper (lower) sign corresponds to the first (second)
qubit. Similarly, the appearance of the variable $\varphi$ in the SCB Josephson
terms introduces an inductive interaction between the SCB and the coupling
junction.

At this point, we are ready to proceed to the circuit Hamiltonian. By
introducing the conjugated variables, $n_i = (1/\hbar)(\partial
L/\partial\dot\phi_{-,i})$, and $n = (1/\hbar)(\partial
L/\partial\dot\varphi)$,
 which have the meaning of dimensionless charges (in the units
of Cooper pairs) on the SCB and on the coupling JJ, respectively, and then
applying the Legendre transformation,
 $H = \sum_i \hbar \,n_i\dot\phi_{-,i} \; + \;\hbar \,n\dot\varphi\; - \;L$,
we get,
\begin{equation}\label{Hclass}
H = \sum_{i=1}^{2} H_{SCB,i} + H_{JJ} + H_{C}.
\end{equation}
Here
\begin{equation}
H_{SCB,i} = E_C\left(n_i-n_{gi}\right)^{2} - 2E_{J}\cos\frac{\varphi}{2}
 \cos \phi_{-,i},
\label{HSCB}
\end{equation}
is the SCB Hamiltonian, where $E_C=(2e)^2/2C_{\Sigma}$, $C_{\Sigma}=2C+C_{g}$
is the total capacitance of the qubit island and $n_{gi}=C_{g}V_{gi}/2e$ is the
(dimensionless) charge induced on the qubit island by the gate voltage. The
JJ Hamiltonian is
\begin{equation}
H_{JJ} = E_C^b \left(n-\frac{n_{g1}-n_{g2}}{2}\right)^2 - E_J^b\cos\varphi
- \frac{\hbar}{2e}I_{b}\varphi, \label{Hosc2class}
\end{equation}
where $E_C^b=(2e)^2/(2C_b+C_{\Sigma})$. The last term in Eq.
(\ref{Hclass}),
\begin{equation}
\fl H_{C} =
\frac{C_g}{C_{\Sigma}}E_C^b\left(n-\frac{n_{g1}-n_{g2}}{2}\right)
\left((n_2-n_{g2})-(n_1-n_{g1})\right) -
\frac{C_g^2}{2C_{\Sigma}^2}E_C^b(n_1-n_{g1})(n_2-n_{g2}),
 \label{H_int_class}
\end{equation}
describes capacitive interaction of the qubits and the JJ, and also direct
qubit-qubit coupling, induced by the gate capacitance.

The Hamiltonian (\ref{Hclass}) is quantized by imposing the canonical
commutation relations, $\left[\phi_{-,j},n_k\right]=i\delta_{jk}$,
$\left[\varphi,n\right]=i$. To incorporate the Coulomb blockade effect, we
take advantage of the periodic SCB potential and impose periodic boundary
conditions on the wave function with respect to the phase $\phi_{-,i}$.
This results in charge quantization on the island. Keeping the system at
low temperature ($k_BT<E_C$) and close to the charge degeneracy point
$n_g=1/2$, restricts the number of excess charges on the island to zero or
one Cooper pair. Assuming the charge regime, $E_C\gg E_J$ for the SCB, and
no transitions to the higher charge states to occur during qubit
operation, we truncate the SCB Hilbert space to these two lowest charge
states. Then the single qubit Hamiltonian reads,
\begin{equation}\label{Hi}
H_i = \frac{E_C}{2}\left(1-2n_{gi}\right)\sigma_{zi} -
E_J\cos\frac{\varphi}{2}\sigma_{xi},
\end{equation}
while the capacitive interaction (\ref{H_int_class}) takes the form,
\begin{equation}
H_{C} = \frac{C_g}{2C_{\Sigma}}E_C^b\left(n-\frac{n_{g1}-n_{g2}}{2}\right)
\left(\sigma_{z2} - \sigma_{z1}\right) -
\frac{C_g^2}{8C_{\Sigma}^2}E_C^b\sigma_{z1}\sigma_{z2}. \label{H_int_q}
\end{equation}
The qubits and JJ interact both capacitively, Eq. (\ref{H_int_q}),  and
inductively (the last term in Eq. (\ref{Hi})). It has been noticed by
Shnirman et al. \cite{Shnirman}, that the capacitive interaction can be fully
transformed into an inductive one. This can be done by using a unitary
rotation conveniently combined with a gauge transformation eliminating the
gate charge-dependent terms in Eqs. (\ref{H_int_q}) and (\ref{Hosc2class})
(this is possible since the charge on the coupling JJ is not quantized). The
corresponding unitary operator is
\begin{equation}\label{Ualpha}
U = \exp\left[ - i\alpha(\sigma_{z2} - \sigma_{z1}) \varphi\right]\,
\exp\left(i{n_{g1}-n_{g2}\over 2}\varphi\right)\,,\qquad \alpha =
{C_g\over 4C_\Sigma}.
\end{equation}
It is straightforward to check that the transformed part of the
Hamiltonian, $U^\dagger (H_{JJ} + H_{C})U$, does not contain any
interactions, while the whole interaction is concentrated in the Josephson
term of the qubit Hamiltonian,
\begin{equation}\label{H_inew}
U^\dagger H_{i}U =
\frac{E_C}{2}\left(1-2n_{gi}\right)\sigma_{zi}
 -
E_J\cos\frac{\varphi}{2}\left[\cos(2\alpha\varphi)\,\sigma_{xi} \pm
\sin(2\alpha\varphi)\,\sigma_{yi}\right].
\end{equation}
The $\alpha$-dependent correction is small when the gate capacitance is
small, $C_g \ll C_\Sigma$.

\subsection{Controllable qubit coupling}
\label{ss2qeff}
Let us consider the main part of the inductive interaction. Making use of
assumption, Eq. (\ref{EjEc}), we consider small phase fluctuations across
the JJ, $\gamma = \varphi - \varphi_0$, around the minimum point
$\varphi_0$, and expand the $\varphi$-dependent terms in Eq.
(\ref{H_inew}) in powers of $\gamma \ll 1$. The quantity $\varphi_0$ is
determined by the applied bias current $I_b$,
\begin{equation}\label{ssp2}
\sin\varphi_0 = {\hbar I_b\over 2eE_J^b}.
\end{equation}
To zeroth order with respect to $\gamma$ we get a free qubit
Hamiltonian which, after additional rotation $U^\prime=\exp\left[
i\alpha(\sigma_{z2} - \sigma_{z1}) \varphi_0\right]$, takes the form,
\begin{equation}\label{Hi2}
H_i  = \frac{E_C}{2}\left(1-2n_{gi}\right)\sigma_{zi} -
E_J\cos\frac{\varphi_0}{2}\sigma_{xi} .
\end{equation}
Controllable qubit-qubit coupling results from the terms which are linear
in $\gamma$ in the expansion of Eq. (\ref{H_inew}). Neglecting the effect
of the small $\alpha$ in these terms (which will be considered in the next
section), we obtain,
\begin{equation}
H_{int} ={1\over 2} E_J\sin\frac{\varphi_0}{2}\ \gamma (\sigma_{x1} +
\sigma_{x2}). \label{H2int}
\end{equation}
It is convenient to combine these terms with the quadratic potential of
the JJ that approximates the tilted Josephson potential near its minimum.
We then get the total Hamiltonian on the form,
\begin{equation}\label{Htot}
H = \sum_i H_i - \lambda
E_J\frac{\sin^2(\varphi_0/2)}{4\cos\varphi_0}\sigma_{x1}\sigma_{x2}\, +\,
H_{osc}
\end{equation}
where $\lambda = E_J/E_J^b$, and
\begin{equation}\label{Hosc}
H_{osc} = E_C^b n^2 + \frac{1}{2}E_J^b \cos\varphi_0\left[\gamma +
\lambda\frac{\sin(\varphi_0/2)}{2\cos\varphi_0}\left(\sigma_{x1}+\sigma_{x2}\right)\right]^2
\end{equation}
is the Hamiltonian of a displaced linear oscillator associated with the
coupling JJ.

The further analysis is significantly simplified if one assumes the
qubit-oscillator interaction to be small, $\lambda \ll 1$, and the oscillator
to be fast on a time scale of qubit evolution, $\omega_b \gg E_J $. In
fact, these assumptions are not needed when the qubits are parked at the
charge degeneracy point, $n_{gi} = 1/2$, because in this case the
interaction term commutes with the qubit Hamiltonians, and the problem is
exactly solvable \cite{Wang}. The assumption on $\lambda$ can be relaxed
for the two-qubit circuit, however for a multi-qubit network it becomes
essential as discussed later.

The imposed constraints together with Eq. (\ref{EjEc}) lead to the
following chain of inequalities:
\begin{equation}\label{constraints}
E_J \ll \omega_b \ll E_J^b.
\end{equation}
Under these constraints, one can neglect the excitation of the oscillator,
which at low temperature will remain in the ground state. This
does not significantly differ from the ground state of the Hamiltonian in
Eq. (\ref{Hosc}). For instance, the estimate for the amplitude of the first
order correction reads,
\begin{equation}
c_{0\rightarrow 1} \sim
\frac{E_{C}(1-2n_{g})}{(E_{J}^b\omega_{b})^{1/2}(\cos\varphi_{0})^{3/4} }
\sim {E_J\over (E_{J}^b\omega_{b})^{1/2}} \ll 1.
\end{equation}
Therefore one can average over the oscillator ground state and drop the
oscillator energy term, because it does not depend on the qubit state
configurations.

Summarizing our derivation, after integrating out the
oscillator, we arrive at the effective two-qubit Hamiltonian,
\begin{equation}\label{Heff}
H_{eff} = \sum_i H_i - \lambda
E_J\frac{\sin^2(\varphi_0/2)}{4\cos\varphi_0}\sigma_{x1}\sigma_{x2} .
\end{equation}
The qubit-qubit coupling term in Eq. (\ref{Heff}) has a clear physical
meaning: it results from interacting persistent currents in the qubit loops.
Indeed, the persistent currents are given in terms of Eqs.
(\ref{LSCB})-(\ref{HSCB}) by the relation $I_i=(2e/\hbar)E_J\sin(\phi_{1i})$,
or identically,
\begin{equation}
I_i =  {2e\over \hbar}E_J \sin{\phi_{+,i}\over 2}\cos\phi_{-,i} = {2e\over
\hbar}{\partial H\over\partial \phi_{+,i} }\,.
\end{equation}
\label{Current}
In the truncated form, this relation reduces to
\begin{equation}\label{}
I_i={e\over\hbar}E_J\sin{\varphi_0\over 2}\,\sigma_{xi}
\end{equation}
 (neglecting phase
fluctuations over the coupling JJ), and the coupling term in Eq.
(\ref{Heff}) can be expressed as an inductive coupling energy of the two
persistent currents, $L_{b}I_1 I_2$, with the Josephson inductance of the
tilted coupling junction,
\begin{equation}\label{Leff2}
L_{b}= {\hbar^2\over 4e^2 E_J^b\cos\varphi_0},
\end{equation}
playing the role of mutual inductance. In the absence of the bias current
when the JJ potential is not tilted, $\sin\varphi_0 = 0$, the persistent
currents are not excited and the coupling is switched off. When the
bias current is applied, the coupling is switched on, and its strength
increases with the bias current because of increasing persistent currents,
and also because of decreasing JJ inductance.

\subsection{Effect of qubit asymmetry}
\label{sass}

Let us consider the effect of small $\alpha$-terms in Eq. (\ref{H_inew}).
Although small, the last term in this equation proportional to $\sigma_y$
leads to an interesting qualitative effect in the qubit coupling, changing
its symmetry. A similar effect is produced by asymmetry of the qubit
junctions. Although an ideal qubit should consist of identical Josephson
junctions, in practice the junction parameters may vary at least within
the range of a few percent. In the asymmetric case, the property of the
symmetric qubit to have zero persistent current when the bias is turned
off is lost. Now a persistent current is spontaneously generated, the
direction of which depends on the charge state of the SCB. This affects the
symmetry of the controllable qubit coupling.

The most important is the variation of the Josephson energy. For an
asymmetric qubit, the Josephson term in Lagrangian, Eq. (\ref{LSCB}), has
the form, $E_{J1}\cos\phi_{1i} + E_{J2}\cos\phi_{2i}$. For small junction
asymmetry, $\delta E_J = E_{J1} - E_{J2} \ll E_{J}$, the Josephson term in
the qubit Hamiltonian, Eq. (\ref{Hi}), acquires the form
\begin{equation}\label{Hiass}
- E_J\cos\frac{\varphi}{2}\sigma_{xi} \pm {\delta E_J \over
2}\sin\frac{\varphi}{2}\sigma_{yi}.
\end{equation}
The second term in this equation, resulting from the junction asymmetry,
has the same $y$-symmetry as the last term in Eq. (\ref{H_inew}).
They can therefore be considered on the same footing
and added to the interaction
Hamiltonian (\ref{H2int}), which now takes the form,
\begin{equation}
\fl H_{int} = \,  E_J \ \gamma \ \left[{1\over
2}\sin\frac{\varphi_0}{2}(\sigma_{x1} + \sigma_{x2}) \right. \left. +
\cos{\varphi_0\over 2}\left(2\alpha - \alpha\varphi_0 \tan{\varphi_0\over
2} - {\delta E_J\over 4E_J}\right)(\sigma_{y2} - \sigma_{y1})\right].
\label{H2int_ass}
\end{equation}
The additional terms give rise to a small direct qubit coupling of the
$xy$-type, in addition to the controllable $xx$-coupling in Eq.
(\ref{Heff}),
\begin{equation}\label{xyint}
{1 \over 4}\lambda E_J\tan\varphi_0\left( 2\alpha- \alpha\varphi_0
\tan\frac{\varphi_0}{2}-{\delta E_J\over
4E_J}\right)(\sigma_{y1}\sigma_{x2} -\sigma_{x1} \sigma_{y2}).
\end{equation}
Although small, this additional coupling term does not commute with the
qubit Hamiltonian even at the degeneracy point, which may complicate the
gate operation discussed towards the end of this paper.

%
\subsection{Residual qubit coupling}
\label{ss2qres}
Even in the absence of bias current, and in the symmetric qubits, there
exist small circulating currents in the qubit loops because of ground
state phase fluctuation in the coupling junction. These currents interact
via the coupling junction, creating a small parasitic coupling of the
qubits. The effect is described by the higher order terms neglected in the
previous discussion. One term, which does not vanish at $\varphi_0=0$ is
due to the interaction via the gate capacitance (cf. Eq.
(\ref{H2int_ass})),
\begin{equation}\label{H2y}
H^{(1)}_{int} = 2\alpha E_J\cos(\varphi_0/2)\,\gamma\,(\sigma_{y2} -
\sigma_{y1}).
\end{equation}
This term is linear in $\gamma$, and it creates a direct parasitic
qubit-qubit coupling via the mechanism discussed in the previous sections,
\begin{equation}\label{Hresy}
H^{(1)}_{res} = 4\alpha^2 \lambda E_J {\cos^2(\varphi_0/2)\over
\cos\varphi_0}\,\sigma_{y1}\sigma_{y2}.
\end{equation}
This coupling is smaller than the controllable coupling by a factor
$\alpha^2=(C_g/4C_\Sigma)^2 \ll 1$.

Obviously, the effect of the junction asymmetry also contributes to this
kind of residual coupling, and can be included in Eq. (\ref{Hresy}), by
making a change, $\alpha \,\rightarrow \, \alpha\, - \,\delta E_J/2E_J$.

%
Another parasitic term is quadratic in $\gamma$,
\begin{eqnarray}
H^{(2)}_{int} = {1\over 8}\lambda E_J^b \cos\frac{\varphi_0}{2}\ \gamma^2
(\sigma_{x1} + \sigma_{x2}). \label{H2x}
\end{eqnarray}
The effect of this interaction is to change the frequency, and hence the
ground state energy of oscillator (\ref{Hosc}), depending on the qubit
state configuration. This squeezing effect creates a direct residual qubit
coupling in the lowest order approximation,
\begin{equation}
\label{H2res}
H^{(2)}_{res} = - \frac{1}{128}\lambda E_J \frac{\hbar\omega_{b}}{E_{J}^b}
\frac{\cos^2(\varphi_0/2)}{(\cos\varphi_0)^{3/2}}\,
\sigma_{x1}\sigma_{x2}.
\end{equation}
This coupling is smaller than the controllable interaction in Eq.
(\ref{Heff}) by a factor, $\hbar\omega_b/E_J^b\ll 1$.

%
\subsection{Maximum coupling strength}
\label{smax}
 Because of the limitation on the gate operation time imposed by
 decoherence, it is desirable that the qubit coupling is as strong as
possible. In our case, the coupling strength is generally determined by the
parameter $\lambda$; the strength however increases with the applied current
bias. This is reflected by a cosine-factor in the denominator in Eq.
(\ref{Heff}), which formally turns to zero at $\varphi_0=\pi/2$. This
corresponds to the point when the bias current approaches the critical
current value for the coupling JJ. At this point the minimum in the tilted
Josephson potential disappears, and the junction switches to the resistive
state, sweeping the qubit phase and thus destroying the qubit. Therefore the
ultimate limitation on the coupling strength is imposed by the switching of
the coupling JJ. The latter may even occur at smaller applied current because
of tunneling through the Josephson potential barrier (macroscopic quantum
tunnelling, MQT). The assumption of a small MQT rate imposes an additional
limitation on the coupling strength to the one imposed by the constraints
(\ref{constraints}). Indeed, because the potential wells of the tilted
Josephson potential become shallow with decreasing Josephson
energy, $E_J^b\cos\varphi_0$, the constraints must be reconsidered,
\begin{equation}\label{constraints2}
E_J \ll \omega_b\sqrt{\cos\varphi_0} \ll E_J^b\cos\varphi_0,
\end{equation}
clearly putting limitations on the maximum allowed tilt.

In order to very roughly estimate an upper bound for the maximum coupling
strength, let us soften requirements (\ref{constraints2}), and consider
the relations

\begin{equation}\label{constraints3}
E_J \sim \omega_b\sqrt{\cos\varphi_0} \sim E_J^b\cos\varphi_0.
\end{equation}
Both the relations can be fulfilled by applying sufficiently large bias
current, and by choosing appropriate plasma frequency. The latter can be
adjusted by shunting the coupling JJ with a large capacitance. The
corresponding relations read,
\begin{equation}\label{constraints4}
\cos\varphi_0 \sim {E_J\over E_J^b},\qquad \omega_b \sim \sqrt{E_J
E_J^b}\,.
\end{equation}
The coupling strength for a tilted JJ is given by the phase-dependent
coupling parameter in Eq. (\ref{Heff}),
\begin{equation}\label{coupling}
\lambda(\varphi_0) = \lambda \,\frac{\sin^2(\varphi_0/2)}{4\cos\varphi_0}\,.
\end{equation}
The maximum value of this parameter is estimated by using Eq.
(\ref{constraints4}),
\begin{equation}\label{maxcoupling1}
\mbox{max} \,\lambda(\varphi_0) \sim 1\,.
\end{equation}
Let us compare this result with the limitation imposed by MQT. For large
applied bias current, the potential well can be approximated with a cubic
curve, and the MQT rate is estimated by\cite{Weiss}
\begin{equation}
\Gamma_{MQT} = \omega_b\sqrt{{30s\over \pi}\cos\varphi_0} \,e^{-s},\qquad
s =
\frac{24E_J^b}{5\omega_b}\frac{(\cos\varphi_0)^{5/2}}{\sin^2\varphi_0}.
\end{equation}
Suppose that the value $\omega_b \sim (1/2)E_J^b(\cos\varphi_0)^{5/2}$ gives
satisfactory small MQT rate ($\sim 10^{-4} E_J$ according to the following
estimates). Under this condition, which is more restrictive that the right
one in Eq. (\ref{constraints3}), the relations in Eq. (\ref{constraints4})
become modified,
\begin{equation}\label{constraints5}
\cos\varphi_0 \sim \left({E_J\over E_J^b}\right)^{1/3},\qquad \omega_b \sim
E_J^{5/6} \left(E_J^b\right)^{1/6}\,,
\end{equation}
leading to a somewhat smaller maximum coupling parameter,
\begin{equation}\label{maxcoupling2}
\mbox{max} \,\lambda(\varphi_0) \sim \left({E_J\over E_J^b}\right)^{2/3} < 1.
\end{equation}
%
%
\subsection{Charge-phase regime}
\label{ss2qchph}
So far, we have assumed the qubit island to be in the charge regime $E_C\gg
E_J$, where the two lowest charge eigenstates, $|n=0\rangle$ and
$|n=1\rangle$, serve as the qubit basis. However, from an experimental point
of view it may be more appealing to work in the charge-phase regime $E_C\sim
E_J$ because the qubit becomes more stable against charge noise when the energy
bands flatten \cite{Vion}. In this regime, the qubit states are given by
Bloch wave functions, consisting of superpositions of many charge states.
Nevertheless, as easily seen, the controllable qubit coupling via a
current biased large JJ will persist also in the charge-phase regime. Indeed,
an essential physical characteristic of the qubit-JJ interaction is the
persistent current in the qubit loop, Eq. (\ref{Current}), $I=(2e/\hbar)
\sin(\phi_+/2)\cos\phi_-$. The magnitude of this current is controlled by the
tilt of the JJ ($\sin(\phi_+/2) = \sin(\varphi_0/2))$, and it is zero when the
JJ is idle, regardless of whether the qubit is in the charge or charge-phase
regime.

Furthermore, an important property of the charge regime is that the qubit-JJ
interaction is diagonal in the qubit eigenbasis when the qubit is parked at the
charge degeneracy point, $n_g=1/2$, i.e. it has $zz$-symmetry in this
eigenbasis. This property simplifies the 2-qubit gate operations discussed
later, and it also allows a quantum non-demolishing measurement of the qubit
by means of current detection using the large JJ,
as discussed in the next section.
We show in this section that this property persists in the charge-phase
regime. Namely, we show that the SCB Hamiltonian truncated to a pair of
lowest Bloch states commutes with the truncated current operator at
$n_g=1/2$. This leads to direct qubit-qubit coupling of $zz$-type in the
qubit eigenbasis.

Let us consider the SCB Hamiltonian $H_{SCB}$, Eq. (\ref{HSCB}), in the charge
basis, $|n\rangle$, and separate the part which does not depend on the gate
charge,
\begin{equation}
H_1 = \sum_{n=-\infty}^{\infty}\Bigg[E_C n(n-1)|n\rangle\langle n|
- \tilde{E}_J\left(|n+1\rangle\langle n|+|n-1\rangle\langle
n|\right)\Bigg],
\end{equation}
from a small part proportional to the departure from the charge degeneracy
point (e.g. during single-qubit manipulation) $\delta n_g(t)=1/2-n_g(t)$,
\begin{equation}\label{H2}
H_2 = \sum_{n=-\infty}^{\infty}2E_C\delta n_{g}(t) n\,|n\rangle\langle
n|,\qquad H_{SCB}=H_1+H_2.
\end{equation}
The notation $\tilde{E}_J=2E_J\cos(\varphi/2)$ is introduced here for
brevity. We split the complete set of the charge eigenstates, $-\infty< n <
\infty$, in the positive and negative charge subsets labelled with $\sigma =
\uparrow,\downarrow$, and $m, \; 1\,<\,m\,<\,\infty$,  such that
\begin{eqnarray}
m = n, & n>0,\nonumber\\
m = 1-n, \qquad& n \leq 0.
\end{eqnarray}
In the basis $|m,\sigma\rangle$, $m=\ldots, 2,1$, the Hamiltonian $H_1$
acquires the form,
\begin{eqnarray}\label{H1}
H_1 &=& \left[
\begin{array}{c|c}
H_0 & -\tilde{E}_JU \\
\hline -\tilde{E}_JU & H_0
\end{array}\right],
\end{eqnarray}
where $H_0$ is tridiagonal, and $U$ contains only a single element,
\begin{equation}
H_0 = \left[
\begin{array}{cccc}
\ddots & \ddots & & \\
\ddots & 6E_C & -\tilde{E}_J & \\
 & -\tilde{E}_J & 2E_C & -\tilde{E}_J \\
 & & -\tilde{E}_J & 0
\end{array}\right], \qquad
\ U = \left[
\begin{array}{ccc}
\ddots & & \vdots\\
 & 0 & 0\\
\ldots & 0 & 1
\end{array}\right].
\end{equation}
A Hadamard rotation, H, in $\sigma$-space,
\begin{equation}\label{Hadamard}
{\textrm H}=\frac{1}{\sqrt 2}\left(\sigma_z+\sigma_x\right),
\end{equation}
takes the basis $|m\uparrow\rangle$, $|m\downarrow\rangle$ to $|m\pm\rangle =
(1/\sqrt{2}) (|m\uparrow\rangle \pm |m\downarrow\rangle)$, and transforms the
matrix in Eq. (\ref{H1}),
\begin{equation}\label{H1transf}
H_1=H_0{\mathbf 1}-\tilde{E}_JU\sigma_x \rightarrow H_0{\mathbf
1}-\tilde{E}_JU\sigma_z.
\end{equation}
Then this matrix acquires a block-diagonal form,
\begin{equation}\label{diagH1}
H_1 = \left[
\begin{array}{cc|cc}
 & & & \\
H_0 -\tilde{E}_JU & & 0 & \\
 & \times & & \\
\hline
 & & & \\
0 & & H_0 +\tilde{E}_JU & \\
 & & & \times
\end{array}\right].
\end{equation}

The above block-diagonal form is suitable for identifying the qubit states in
the charge-phase regime. Indeed, when the Josephson energy is tuned to zero,
the lower-corner elements of the blocks, marked with $\times$, correspond to
the lowest energy states of the SCB. In fact, the Hadamard transformation
(\ref{Hadamard}) corresponds to the rotation to the eigenbasis of a charge
qubit at the charge degeneracy point, which coincides with the current
eigenbasis. When the Josephson energy increases, the eigenstates of the
matrix (\ref{diagH1}) become superpositions of many charge states, which
however does not mix the charge superpositions denoted with indices $+$ and
$-$, and can be obtained by independent rotations of the matrix blocks.
During these rotations, although the two lower-corner eigenvalues
marked with $\times$ do
change, they however remain the lowest energy levels. This follows from the
fact that the eigenvalues of the Mathieu equation do not cross when the
amplitude of the potential increases \cite{BenderOrszag}. Therefore the
charge qubit eigenstates develop to the lowest energy Bloch states, which are
identified as the charge-phase qubit eigenbasis $|E_+\rangle$ and
$|E_-\rangle$.

Let us evaluate the form of the current operator, Eq. (\ref{Current}), in the
charge-phase qubit eigenbasis. The current operator in the charge
representation is proportional to the operator $X=|n+1\rangle\langle
n|+|n-1\rangle\langle n|$. In the $|m\sigma\rangle$-basis, this operator is
written as
\begin{equation}
X = X_0{\mathbf 1} + U\sigma_x,
\end{equation}
where,
\begin{equation}
X_0 = \left[
\begin{array}{cccc}
 & \ddots & & \\
\ddots & & 1 & \\
 & 1 & 0 & 1 \\
 & & 1 & 0
\end{array}\right].
\end{equation}
In the eigenbasis of $H_1$, i.e. after the Hadamard transformation, this
operator acquires a block-diagonal form,
\begin{equation}
X \rightarrow \left[
\begin{array}{c|c}
X_0+U & 0 \\
\hline 0 & X_0-U
\end{array}\right],
\end{equation}
which means that $X$ does not couple the states $|E_+\rangle$ and
$|E_-\rangle$.

One implication of this result is that the qubit-qubit coupling in
the charge-phase regime will still be of $zz$-type in the
truncated Hilbert space, i.e. diagonal in the qubit eigenbasis. It also
means that current measurement will not mix the qubit states, i.e.,
current detection provides a means for quantum non-demolition
measurements.

Finally we analyze the term $H_2$ in Eq. (\ref{H2}), which is non-zero only
when the gate charge deviates from the degeneracy point. In the
$|m\sigma\rangle$ representation this term has form,
\begin{eqnarray}
H_2 &=& 2E_C\delta n_{g}(t) \left[
\begin{array}{c|c}
D_{\uparrow} & 0 \\
\hline 0 & D_{\downarrow}
\end{array}\right],
\end{eqnarray}
where $D_{\uparrow}$ and $D_{\downarrow}$ are diagonal matrices,
\begin{eqnarray}
D_{\uparrow} = {\textrm{diag}}\left(\dots,3,2,1\right)\nonumber\\
D_{\downarrow} = {\textrm{diag}}\left(\dots,-2,-1,0\right).
\end{eqnarray}
After the Hadamard rotation it acquires the form,
\begin{equation}
H_2 \rightarrow E_C\delta n_g(t)\left[
\begin{array}{c|c}
{\mathbf 1} & D_{\uparrow}-D_{\downarrow} \\
\hline D_{\uparrow}-D_{\downarrow} & {\mathbf 1}
\end{array}\right].
\end{equation}
Thus after truncation to the qubit basis, $D_{\uparrow}-D_{\downarrow}$
provide off-diagonal elements, which couple the qubit states and can be
employed for qubit manipulation.

\section{Coupling via read-out junctions}
\label{s2qm}
The readout circuit is an important ingredient of the qubit network, which
must be explicitly included in the consideration. We discuss here the
readout method successfully tested on a single qubit by the Saclay group
\cite{Vion}. With this method, the persistent current flowing in the qubit
loop is excited and measured by using a large Josephson junction in the
qubit loop, as shown in Fig. \ref{2qubm}. To do the measurement, a large
dc current is applied to the junction so that the net current through the
junction either exceeds the critical value or not depending on the
direction of the persistent current in the loop. In the former case, the
measurement JJ switches to a resistive state, which is detected by
measuring a dc voltage across the JJ; in the latter case, no voltage is
detected. This method of threshold detection is quite invasive, sweeping
the phase across the qubit and creating a large number of quasiparticles.
A recently tested more gentle method \cite{Siddiqi} utilizes an ac driving
current with comparatively small amplitude applied to the JJ, and measures
the qubit-state dependent ac response.

We now analyze the compatibility of such measurement methods, via a large
JJ, with our coupling scheme. Before proceeding with the calculations we
note that one may distinguish two cases: measurement and coupling. In the
measurement case, the bias current is applied only to one single
measurement junction. This will excite the persistent current in the
corresponding qubit loop, allowing qubit readout, while the neighboring
qubit loop will remain, as we will see, in the idle state, neglecting the
effect of a small parasitic coupling, and this qubit will not be
destroyed. In the coupling case, bias current sent through both
measurement junctions in Fig. \ref{2qubm} will create persistent currents
in both qubit loops, resulting in the qubit coupling discussed above.
This physical picture implies that qubit-qubit coupling can be achieved
even without sending current through the coupling junction \cite{Lantz}.

\subsection{Measurement of individual qubits}
\label{ss2qmmeas}
\begin{figure}[hbt]
\begin{center}
\includegraphics[width=5cm]{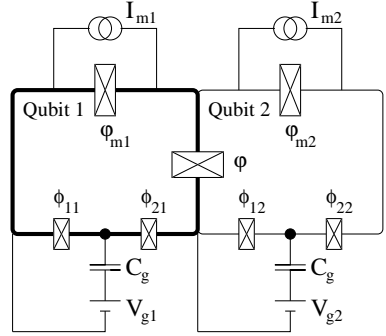}
\caption{Applied currents $I_{mi}$ across large measurement JJs can be used
for reading out the current state of qubit $i$ and for qubit coupling.}
\label{2qubm}
\end{center}
\end{figure}
When a measurement JJ is included in each qubit loop (Fig. \ref{2qubm}),
the corresponding terms must be added to the circuit Lagrangian Eq.
(\ref{Lorig}):
\begin{equation}\label{Lm}
L_{mi} = \frac{\hbar^2C_{mi}}{2(2e)^2}\dot{\varphi}_{mi}^{2} + E_{Ji}^m\cos
\varphi_{mi} + \frac{\hbar}{2e}I_{mi}\varphi_{mi}.
\end{equation}
 Here
$\varphi_{mi}$ denotes the phase across the measurement junction of the
$i$-th qubit, and $I_{mi}$ is the applied current. The phase quantization
relations (\ref{phases}) will now change,
\begin{equation}\label{phases2}
\phi_{+,1}+\varphi - \varphi_{m1}=0, \qquad \phi_{+,2}-\varphi - \varphi_{m2}
=0\,,
\end{equation}
giving rise to interaction of the qubits with the measurement JJ, in addition
to the coupling JJ, in the charge sector as well as in the current sector.
As before, it
is possible, to eliminate the capacitive interaction, but now it
is more convenient to do it on the Lagrangian level. The interaction
via the gate
capacitance is eliminated via transformation of the qubit variable,
\begin{equation}\label{alpha}
\phi_{-,i} \; \rightarrow \; \phi_{-,i} + 2\alpha(\varphi_{mi} \mp \varphi),
\end{equation}
which is equivalent to the transformation in Eq. (\ref{Ualpha}) (the upper
(lower) sign corresponds to the first (second) qubit). As already
described, this interaction leads to a small residual direct $yy$ qubit
coupling, which we will omit from the further discussion. Similarly, the
capacitive interaction via the qubit capacitance $C$ can be eliminated by
transformation of the measurement JJ variable,
\begin{equation}\label{beta}
\varphi_{mi} \; \rightarrow \; \varphi_{mi} \pm \beta \varphi, \qquad \beta =
{C\over 2C_{mi} + C}.
\end{equation}
This transformation will only slightly affect the inductive interaction since
$\beta$ is small.

At this point, we proceed to the quantum description of the circuit and
truncate the qubit Hamiltonian, assuming the charge regime, $E_C \gg E_J$.
The circuit Hamiltonian will take the form,
\begin{equation}\label{HimiJJ}
H = \sum_{i=1}^{2} (H_{i} + H_{mi} ) + H_{JJ}  ,
\end{equation}
where
\begin{equation}\label{Him}
H_i = \frac{E_C}{2}\left(1-2n_{gi}\right)\sigma_{zi} -
E_J\cos\frac{(1-\beta)\varphi \mp \varphi_{mi}}{2}\;\sigma_{xi},
\end{equation}
refers to the qubits, while
\begin{equation}
H_{mi} = E_{Ci}^m n_{mi}^2 - E_{Ji}^m\cos(\varphi_{mi} \pm \beta\varphi) -
\frac{\hbar}{2e}I_{mi}(\varphi_{mi} \pm \beta\varphi), \label{HmJJ}
\end{equation}
and
\begin{equation}\label{HJJ2}
H_{JJ} = E_C^b n^2 - E_J^b\cos\varphi,
\end{equation}
are the Hamiltonians of the measurement JJ and the coupling JJ,
respectively. In these equations, $E_{Ci}^m = (2e)^2/(2C_{mi}+C)$, and
$E_C^b = (2e)^2/2C_{b\Sigma}$, $C_{b\Sigma} = C_b +
\sum_i[C(C_{mi}+C)/(2C_{mi}+C)]$; the current applied to the coupling junction
is absent because we will focus on the effect of the measurement
junctions.

In the measurement regime only a single external current, say $I_{m1}$, is
applied. The steady state point for the 3-JJ network, ($\varphi_0$,
$\varphi_{mi,0}$), is found from Eqs. (\ref{Him})-(\ref{HJJ2}) in the main
approximation with respect to the small parameters $\beta \approx C/C_b$ and
$\lambda = E_J/E_J^b$,
\begin{eqnarray}\label{steady}
\sin\varphi_{m1,0} = {\hbar\over 2e}{I_{m1}\over E_{J1}^m},\nonumber\\
\varphi_0 = {\lambda \over 2}\sin{\varphi_{m1,0}\over 2}\,\sigma_{x1},
\nonumber\\
\varphi_{m2,0} = \beta\varphi_0 - {\lambda \over 2}\sin{\varphi_0\over
2}\,\sigma_{x2}.
\end{eqnarray}
It follows from these equations that indeed the phases across the coupling JJ
and the second measurement JJ remain negligibly small even though the first
measurement junction may be biased at the critical level. Thus, the
constraint (\ref{constraints}) is essential for not disturbing the other
qubit while the first qubit is measured.
%
\subsection{Qubit coupling via readout junctions}
\label{ss2qmeff}
\label{ss2qmL}

 In the case of qubit-qubit coupling, both measurement junctions
are biased, while the coupling junction is not tilted by external bias,
\begin{equation}\label{ssp2m}
\sin\varphi_{mi,0} = {\hbar\over 2e}{I_{mi}\over E_{Ji}^m},\qquad \varphi_0 =
0.
\end{equation}
Expanding the potential terms in Eqs. (\ref{Him})-(\ref{HJJ2}) around the
steady state point up to second order with respect to small phase
fluctuations, $\gamma$ and $\gamma_{mi} = \varphi_{mi} - \varphi_{mi,0}$, and
neglecting $\beta$-corrections, we may present the Hamiltonian on the form
\begin{equation}\label{Hm}
H = \sum_{i=1}^{2} H_{i} + H_{osc}  + H_{int}.
\end{equation}
Here
\begin{equation}\label{Him2}
H_i = \frac{E_C}{2}\left(1-2n_{gi}\right)\sigma_{zi} - E_J\cos
{\varphi_{mi,0}\over 2}\;\sigma_{xi},
\end{equation}
is the qubit Hamiltonian, which differs from the one in Eq. (\ref{Hi}) by
the phase of the measurement JJ  substituting for the phase of the
coupling JJ. The next term,
\begin{equation}\label{oscm}
\fl
H_{osc} = \,E_C^b n^2 + E_{C1}^m n_{m1}^2 + E_{C2}^m n_{m2}^2
 +  {1\over 2} (E_J^b\gamma^2 +
E_{J1}^m\cos\varphi_{m1,0}\gamma_{m1}^2 +
E_{J2}^m\cos\varphi_{m2,0}\gamma_{m2}^2 \,)
\end{equation}
describes uncoupled linear oscillators, while the interaction is described
by the last term,
\begin{eqnarray}\label{intm}
\fl
H_{int} = \frac{\lambda E_J^b}{2}
\left[(\Delta_1 + \Delta_2)\ \gamma^2
+\Delta_1\gamma_{m1}^2 + \Delta_2\gamma_{m2}^2 \right]
+ \lambda E_J^b \left(\Delta_2\gamma_{m2} - \Delta_1\gamma_{m1}\right)
\gamma\nonumber\\
+ \lambda E_J^b \left[(B_1 - B_2)\gamma - B_1\gamma_{m1} -
B_2\gamma_{m2}\right]. \label{Hbeforeoscm}
\end{eqnarray}
Here we introduced for brevity the following notations,
\begin{eqnarray}\label{DeltaB}
\Delta_i = (1/4)\cos(\varphi_{mi,0}/2)\sigma_{xi}, \nonumber\\
  B_i = -(1/2)\sin(\varphi_{mi,0}/2)\sigma_{xi}.
\end{eqnarray}

Now our goal will be to eliminate the linear terms in gammas in
Eq. (\ref{intm}),
which can be easily done by using oscillator normal modes. To this end we
rewrite the potential part of Eqs. (\ref{oscm}), (\ref{intm}) in a symbolic
form in terms of a 3-vector $\hat\gamma = (\,\gamma, \, \gamma_{m1}, \,
\gamma_{m2}\,)$,
\begin{equation}\label{intm2}
 {1\over 2}E_J^b\,
\hat\gamma(\hat D + \lambda\hat\Delta) \hat\gamma + \lambda E_J^b \hat B
\hat\gamma.
\end{equation}
Here $\hat D$ is a diagonal matrix representing the free oscillator
potentials in Eq. (\ref{oscm}), while the $3\times3$ matrix $\hat\Delta$
and the 3-vector $\hat B$ represent the interaction in Eqs. (\ref{intm})
and (\ref{DeltaB}). Without loss of generality we may assume the charging
energies of the oscillators to be equal \cite{comment_rescale}. Then
performing rotation to the eigenbasis $\hat\gamma^\prime$ of the matrix
$\hat D + \lambda\hat\Delta$ and then shifting the variable, $\tilde\gamma
= \hat\gamma^\prime + \lambda \hat D^{\prime-1}\hat B^\prime$ (here the
prime indicates a new basis), we get,
\begin{equation}\label{intm3}
 {1\over 2}E_J^b\,
\tilde\gamma\hat D^\prime \tilde\gamma - {\lambda^2\over 2}E_J^b \hat B
(\hat D + \lambda\hat\Delta)^{-1} \hat B.
\end{equation}
The last term in this equation, which is conveniently written in the
original basis, gives a direct controllable qubit-qubit coupling
similar to the one in Eq. (\ref{Heff}),
\begin{equation}\label{H_int_m}
H_{int} = {1\over 4}\lambda E_{J}
\sin\frac{\varphi_{m1,0}}{2}\sin\frac{\varphi_{m2,0}}{2}\
\sigma_{x1}\sigma_{x2}.
\end{equation}
As expected, this coupling is switched off when one or both
measurement junctions are idle, and it is switched on only when both the
measurement junctions are biased. We emphasize that this coupling does not
require biasing of the coupling junction.

The first term in Eq. (\ref{intm3}) gives, after averaging over the
oscillator ground state, the oscillator ground state energy,
$(\hbar\omega_b/2){\mbox {Tr}} \sqrt{\hat D^\prime}$ (remember that $\hat D'$
is diagonal). Treating $\lambda\hat\Delta$ as a small perturbation, we find
the first perturbative correction to the matrix spectrum, $\hat D^\prime \,=
\,\hat D + \lambda\,{\mbox {diag}} \hat\Delta$. It is easy to see that only
the contribution of the coupling JJ contains the dependence on the qubit
state configuration. The relevant matrix element has the explicit form
$(1/2)E_J^b\,[1 \,+\, \lambda (\Delta_1 + \Delta_2)]$, and yields the
residual interaction
\begin{equation}
H_{ires} = - \frac{1}{128}\lambda E_J \frac{\hbar\omega_{b}}{E_{J}^b}
\cos{\varphi_{m1,0}\over2}\cos{\varphi_{m2,0}\over2} \,
\sigma_{x1}\sigma_{x2}.
\end{equation}
This is a small residual interaction substituting for Eq. (\ref{H2res}) in the
present case.

\section{Multiqubit network}
\label{sNq}
\begin{figure}[hbt]
\begin{center}
\includegraphics[width=0.6\textwidth]{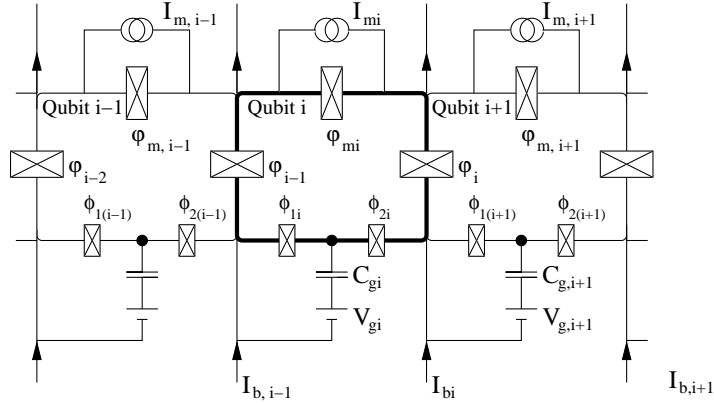}
\caption{A system of $N$ coupled charge qubits. $V_{gi}$ controls individual
qubit, whereas $I_{bi}$ controls the coupling of qubits $i$ and $i+1$.
$I_{mi}$ can be used to read out the current state of qubit $i$ and also for
qubit coupling. There are no coupling JJs at the ends,
$\varphi_0=\varphi_N=0$. } \label{Nqub}
\end{center}
\end{figure}

To implement useful quantum algorithms, controllable systems with large
numbers of qubits are needed. In this section we will show that the
effective qubit-qubit coupling derived in sections \ref{s2q} and
\ref{s2qm} can be generalized to a chain of $N$ qubits with each qubit
being coupled to its nearest neighbors via current-biased JJs and each
having its own read-out device, as shown on Fig. \ref{Nqub}.
\subsection{Circuit Hamiltonian}
\label{ssNqL}
The Lagrangian of the $N$-qubit circuit presented in Fig. \ref{Nqub} can
be written as a straightforward generalization of Eqs. (\ref{Lorig}),
(\ref{LSCB}), (\ref{LJJ}), and (\ref{Lm}),
\begin{equation}
\label{multiL} L = \sum_{i=1}^N \left[L_{SCB,i} + L_{mi}\right] +
\sum_{i=1}^{N-1} L_{JJ,i} \,.
\end{equation}
Since now there are two coupling JJs per qubit loop, the flux
quantization relation (\ref{phases2}) must be extended,
\begin{equation}
\phi_{+,i}+\varphi_i-\phi_{mi}-\varphi_{i-1}=0,
\end{equation}
leading to a more complex form of the interaction among the qubits and the
coupling and measurement JJs. Nevertheless, our previous strategy for
elimination of the interaction in the charge sector still works. Generalizing
Eq. (\ref{alpha}),
\begin{equation}\label{alphaN}
\phi_{-,i} \; \rightarrow \; \phi_{-,i} + 2\alpha(\varphi_{mi}+
\varphi_{i-1}+\varphi_i).
\end{equation}
allows us to decouple qubit charges; the resulting weak interaction in the
current sector yields a direct parasitic $yy$ qubit coupling, similar to the
one in Eq. (\ref{Hresy}). Further transformation, generalizing Eq.
(\ref{beta}),
\begin{equation}\label{betaN}
\varphi_{mi} \; \rightarrow \; \varphi_{mi} + \beta
(\varphi_i-\varphi_{i-1}), \;\;\; \beta = {C\over 2C_{mi} + C},
\end{equation}
decouples the charges of the measurement JJs, and yields weak additional
interaction in the current sector, which will be also omitted.

After the transformations (\ref{alphaN}) and (\ref{betaN}), the only
capacitive interaction which remains in the Lagrangian (\ref{multiL}) is
the interaction among the coupling JJs. This interaction is proportional
to a small qubit capacitance $C$, while the diagonal terms are
proportional to much larger capacitances of the coupling JJs, $C_{bi} \gg
C$. We assume that the JJ capacitances are different so that $C_{bi} -
C_{bj} \gg C$; this is a realistic assumption because a spread of the
junction characteristics during fabrication usually exceeds 10\%. Under
this assumption, the diagonalization of the capacitance matrix will
introduce small corrections to the inductive interaction, corrections which
will not provide any qualitative changes and can be omitted.

With diagonal kinetic terms in the Lagrangian (\ref{multiL}),
it is straightforward to
proceed to the truncation of the quantum Hamiltonian,
\begin{equation}\label{HN}
H = \sum_{i=1}^N \left[H_i + H_{mi}\right] + \sum_{i=1}^{N-1}H_{JJ,i}.
\end{equation}
In the charge regime, $E_J\ll E_C$, the qubit Hamiltonian will take the form,
\begin{equation}
H_i = \frac{E_{C}}{2}(1-2n_{gi}) \sigma_{zi} -
E_J\cos\frac{\varphi_i-\varphi_{mi}-\varphi_{i-1}}{2}\sigma_{xi},
\label{HN1}
\end{equation}
while the large-JJ terms will not change,
\begin{eqnarray}
H_{mi} = E_{Ci}^m n_{mi}^2 - E_{Ji}^m\cos\varphi_{mi} -
\frac{\hbar}{2e}I_{mi} \varphi_{mi},\nonumber\\
H_{JJ,i} = E_C^b n_i^2 - E_J^b\cos\varphi_i -
\frac{\hbar}{2e}I_{bi}\varphi_i.
\label{HjjN}
\end{eqnarray}

In these equations, the effective capacitances of the measurement JJs are
the same as in Eq. (\ref{Hm}), while the effective capacitances of the
coupling JJs are straightforward generalizations of the one in
Eq. (\ref{HJJ2}), namely
$C_{b\Sigma, i}=C_b+C[(C_{mi}+C) / (2C_{mi}+C) +
(C_{m,i+1}+C) / (2C_{m,i+1}+C)]$.

\subsection{Direct qubit-qubit coupling}
\label{ssNqcint}

The next step in the derivation of the direct qubit-qubit coupling is to
eliminate the large JJs, following the previous procedure for the
two-qubit case. After expanding the Hamiltonian,
Eqs.(\ref{HN})-(\ref{HjjN}), with respect to small fluctuations around
the steady
state points, $(\varphi_{i0},\varphi_{mi,0})$, determined by the applied
controlling and measurement currents,
\begin{equation}\label{sspN}
\sin\varphi_{mi,0} = {\hbar\over 2e}{I_{mi}\over E_{Ji}^m},\qquad
\sin\varphi_{i0} = {\hbar\over 2e}{I_{bi}\over E_{J}^b}\,,
\end{equation}
we get the qubit terms (\ref{HN1}) with steady state phases,
(\ref{sspN}), in the Josephson terms. The qubits interact with a
subnetwork of the linear oscillators,
\begin{equation}
H_{osc} = \sum_{i=1}^{N-1}\left[E_C^b
n_i^2+\frac{E_J^b}{2}\cos\varphi_{i0}
\gamma_i^2\right]
+ \sum_{i=1}^N\left[E_{Ci}^m n_{mi}^2+\frac{E_J^b}{2}\cos\phi_{mi0}
\gamma_{mi}^2\right], \label{HfreeoscN}
\end{equation}
via the interaction Hamiltonian, which also connects the oscillators,
\begin{eqnarray}
\label{HclassN}
\fl
H_{int} = \frac{\lambda E_J^b}{2}\sum_{i=1}^{N-1}\left[\left(
\Delta_i+ \Delta_{i+1}\right)\gamma_i^2-2\Delta_i\gamma_i\gamma_{i-1}\right]
\nonumber\\
+ \frac{\lambda E_J^b}{2}\left[\sum_{i=1}^N \Delta_i\gamma_{mi}^2 +
2\sum_{i=1}^{N-1}\left(\Delta_i\gamma_i\gamma_{mi}-\Delta_{i+1}
\gamma_i\gamma_{m,i+1}\right)\right]\nonumber\\
+ \lambda E_J^b \left[\sum_{i=1}^{N-1}(B_i-B_{i+1}) \gamma_i-
\sum_{i=1}^N B_i\gamma_{mi} \right]
\end{eqnarray}
The quantities $\Delta$ and $B$ now contain also the phases of the two
coupling JJs (cf. Eq. (\ref{DeltaB})),
\begin{eqnarray}\label{DeltaBN}
\Delta_i =
\frac{1}{4}\cos\frac{\varphi_{i0}-\phi_{mi0}-\varphi_{i-1,0}}{2}
\sigma_{xi},\nonumber\\
B_i =
\frac{1}{2}\sin\frac{\varphi_{i0}-\phi_{mi0}-\varphi_{i-1,0}}2\,\sigma_{xi}.
\end{eqnarray}

The interaction (\ref{HclassN}) can be presented in the symbolic form of
Eq. (\ref{intm2}) by introducing the $2N-1$-vector
$\hat\gamma=\left(\gamma_1,\gamma_2,\ldots,\gamma_{m1},\gamma_{m2},\ldots
\right)$, the $2N-1\times 2N-1$ matrix $\hat\Delta$ representing the
oscillator interaction, and the $2N-1$-vector $\hat B$ representing the
qubit-oscillator interaction. Then we proceed to Eq. (\ref{intm3}) by
performing the diagonalization, and shifting the oscillator variables as
described after Eq. (\ref{intm2}). The result of
this procedure is as follows \cite{Lantz}: Assuming no applied measurement
currents, the coupling induced by only tilting coupling JJs has the
form,
\begin{equation}
H_{int} = \sum_{i=1}^{N-1}\frac{\lambda E_J}{4\cos\varphi_{i0}}
\sin\frac{\varphi_{i0}-\varphi_{i-1,0}}{2}
\sin\frac{\varphi_{i+1,0}-\varphi_{i0}}{2}\ \sigma_{xi}\sigma_{x,i+1}.
\label{HintN2}
\end{equation}
On the other hand, when the coupling JJs are kept idle while the
measurement junctions are biased, the coupling has the form,
\begin{equation}
H_{int} = \sum_{i=1}^{N-1}\frac{\lambda E_J}{4} \sin\frac{\phi_{mi0}}{2}
\sin\frac{\phi_{m,i+1,0}}{2}\ \sigma_{xi}\sigma_{x,i+1}, \label{HintN1}
\end{equation}
Whichever way the coupling is initiated, one is allowed to simultaneously
perform a number
of two-qubit gates on different qubit pairs, as long as the
qubit pairs are separated by at least one idle qubit. The small residual
$xx$ coupling resulting from the shift of the oscillators ground
energy is restricted to the neighboring qubits and given by Eq.
(\ref{H2res}).

We conclude this section with a discussion of the effect of different
Josephson energies of the qubit junctions, $E_{Ji}$, and the coupling JJ,
$E^b_{Ji}$. This variation can easily be taken into account by introducing
numerical scaling factors, $E_{Ji}= \xi_iE_{J}$, and $E^b_{Ji}=
\xi^b_iE^b_{J}$. Then, while deriving Eq. (\ref{HclassN}), these scaling
factors can be included in the definition of the quantities $\Delta_i$ and
$B_i$ in Eq. (\ref{DeltaBN}). As a result, the coupling energies $\lambda
E_J$ in the final results, Eqs. (\ref{HintN2}), and (\ref{HintN1}), are
replaced by
\begin{equation}\label{}
\lambda E_J  = {E_J^2\over E_J^b} \rightarrow {E_{Ji}E_{Ji+1}\over
E_{Ji}^b}.
\end{equation}

\section{Gate operations with the qubit network}
\label{suse}
All quantum algorithms can be implemented using a limited
universal set of gates.
One such set consists of the controlled-NOT (CNOT) gate together with
 single qubit gates
\cite{Barenco}. In this section, we will describe how to perform a CNOT gate on
two neighbouring qubits in the above mentioned charge qubit network.
Using a sequence of two-qubit operations on nearest neighbours only,
two-qubit operations on arbitrary qubits in the chain can be performed
\cite{Schuch}. The CNOT gate presented here is composed of a control-phase
(CPHASE) gate and two kinds of single qubit gates, a phase gate and the
Hadamard gate.

By default, during qubit operations the qubits are parked at the charge
degeneracy point, where they are more stable against charge noise
\cite{Vion,Duty} and the qubit levels are maximally separated from
higher states.
The computational basis is chosen to be the current basis, which is the
eigenbasis at the charge degeneracy point and differs from the charge basis
by the rotation $\sigma_x\leftrightarrow\sigma_z$.

\subsection{Single qubit operations}
\label{ssuses}
When $I_{bi}=I_{b,i-1}=I_{mi}=0$, qubit $i$ is disconnected from the network
to first order
and single qubit gates can be performed. The time evolution is determined by
the single qubit Hamiltonian $H_i$, Eq. (\ref{Hi2}) or (\ref{Him2}),
\begin{equation}
H_{i}=  E_C\ \delta n_{gi}(t)
 \sigma_{xi} - E_J\sigma_{zi},
\end{equation}
here $\delta n_{gi}(t)=1/2-n_{gi}(t)$ is the deviation from the charge
degeneracy point.

In the idle state, the non-zero energy level splitting results in a phase gate
${\textrm S}_{\theta}$ being performed on the qubit;
\begin{equation}
{\textrm S}_{\theta}=\left\{
\begin{array}{ccc}
|0\rangle & \rightarrow & e^{i\theta/2}|0\rangle \\
|1\rangle & \rightarrow & e^{-i\theta/2}|1\rangle
\end{array}\right.,
\end{equation}
where $\theta$ depends on the elapsed time $T$ through $T=\theta/(2E_J)$.

A particularly useful case is the $S_{3\pi/2}$-gate which will be referred to
as the Z-gate,
\begin{equation}
\label{zgate}
{\textrm Z}=\left\{
\begin{array}{ccc}
|0\rangle & \rightarrow & |0\rangle \\
|1\rangle & \rightarrow & i|1\rangle
\end{array}\right. .
\end{equation}

Another useful single qubit operation is the Hadamard gate H,
\begin{equation}
\label{Hgate}
\textrm{H} =\left\{
\begin{array}{ccc}
|0\rangle & \rightarrow & \frac{1}{\sqrt{2}}
\left(|0\rangle + |1\rangle \right) \\
|1\rangle & \rightarrow & \frac{1}{\sqrt{2}}\left(|0\rangle - |1\rangle \right)
\end{array}\right. ,
\end{equation}
which can be implemented by applying a microwave pulse at the gate
\cite{NakamuraPRL,Vion}, $\delta n_{gi}(t)=A\cos(2E_J t)$,
during a time $T=\pi/2A$. Choosing the amplitude $A$ involves a trade-off
between keeping the operation time short and minimizing the deviations
from the charge degeneracy point.

\subsection{Two-qubit gates}
\label{ssusem}
\begin{figure}[hbt]
\begin{center}
\includegraphics[width=0.6\textwidth]{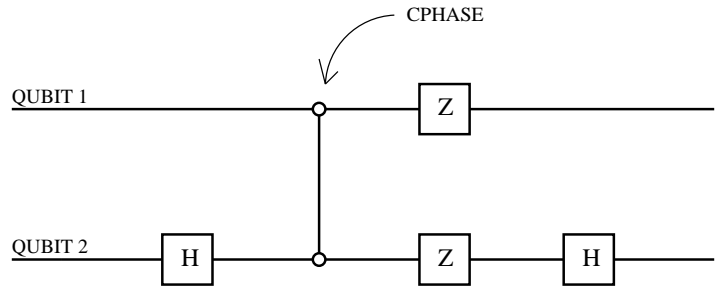}
\caption{A CNOT operation using single qubit gates (Hadamard and phase
gates) and CPHASE. Time runs from left to right.} \label{cphasefig}
\end{center}
\end{figure}

A two-qubit gate involving qubits $i$ and $i+1$ is created by applying a
bias current $I_{bi}$ at the intersection between the two qubits, or by
simultaneously applying measurement currents $I_{mi}$ and
$I_{m,i+1}$. The qubits are coupled according to the coupling terms Eqs.
(\ref{HintN2}) and (\ref{HintN1}), while their individual time evolutions
are determined by $H_i$, Eq. (\ref{Hi2}) or (\ref{Him2}). As an example,
when applying the bias current $I_{bi}$, the Hamiltonian of the two
interacting qubits reads,
\begin{equation}
\fl
 H_i + H_{i+1} + H_{int}^{(1)}
 = -E_J\cos\frac{\varphi_{i0}}{2}(\sigma_{zi} + \sigma_{z,i+1}) -
\frac{\lambda E_J}{4\cos\varphi_{i0}}\sin^2{\varphi_{i0} \over 2}
\sigma_{zi}\sigma_{z,i+1}.
\end{equation}
Choosing operation time and bias current amplitude properly results in the
entangling control-phase (CPHASE) gate,
\begin{equation}
\textrm {CPHASE}=\left\{
\begin{array}{ccc}
|11\rangle & \rightarrow & i|11\rangle\\
|10\rangle & \rightarrow & |10\rangle\\
|01\rangle & \rightarrow & |01\rangle\\
|00\rangle & \rightarrow & i|00\rangle
\end{array}
\right. .
\label{cphaseeq}
\end{equation}
Moreover, a CNOT gate is created by combining CPHASE with single qubit
gates such as the Z-gate, Eq. (\ref{zgate}),
 and the Hadamard gate, Eq. (\ref{Hgate}), as shown in Fig. \ref{cphasefig}.
Thus it is possible to perform a universal set of quantum gates,
 and therefore any quantum algorithm, with the investigated charge-qubit
network.

\ack

The authors acknowledge useful discussions with D. Esteve and Yu. Makhlin.
The support from EU-SQUBIT2 and Swedish grant agencies SSF and VR are
gratefully acknowledged.


%

\begin{thebibliography}{999}
\bibitem{comment_rescale}
By rescaling $\gamma$-variables, the charging energies can always be made
equal; then rescaling constant factors will appear in $\Delta_i$ and
$B_i$, introducing no qualitative changes.
\end{thebibliography}
\end{document}